\RequirePackage{fix-cm}
\documentclass[smallextended]{svjour3}       % onecolumn (second format)
\smartqed  % flush right qed marks, e.g. at end of proof
\usepackage{graphicx}
\usepackage{color}

\def \ee {\begin{equation}}
\def \eee {\end{equation}}
\def \eqe {\begin{eqnarray}}
\def \eqee {\end{eqnarray}}
\definecolor{hotmagenta}{rgb}{1.0, 0.11, 0.81}

\usepackage[round]{natbib}   % omit 'round' option if you prefer square brackets
\begin{document}

\title{Cellular switches orchestrate rhythmic circuits}


\author{Guillaume Drion, Alessio Franci, Rodolphe Sepulchre}


\institute{Guillaume Drion \at
		Department of Electrical Engineering and Computer Science, University of Liege, Liege, Belgium.\\
              \email{gdrion@uliege.be}           %  \\
           \and
           Alessio Franci \at
           National Autonomous University of Mexico, Science Faculty, Department of Mathematics, Coyoac\'{a}n, D.F. M\'{e}xico.\\
              \email{afranci@ciencias.unam.mx}
           \and
           Rodolphe Sepulchre \at
           Department of Engineering, University of Cambridge, Cambridge, United Kingdom\\
              \email{r.sepulchre@eng.cam.ac.uk} \\
              \newline
              GD and AF contributed equally to this work.
}

\date{Received: date / Accepted: date}

\maketitle

\begin{abstract}
Small inhibitory neuronal circuits have long been identified as key neuronal motifs to generate and modulate
the coexisting rhythms of various motor functions. Our paper highlights the role of a cellular switching mechanism to orchestrate such circuits. The cellular switch makes
the circuits reconfigurable, robust, adaptable, and externally controllable. Without this cellular mechanism, the circuits rhythms
entirely rely on specific tunings of the synaptic connectivity, which makes them rigid, fragile, and difficult to control externally.
We illustrate those properties on the much studied architecture of a small network controlling both the pyloric and gastric rhythms of crabs.
The cellular switch is provided by a slow negative conductance often neglected in  mathematical modeling of central pattern generators. We propose
that this conductance is simple to model and key to computational  studies of rhythmic circuit neuromodulation.

\end{abstract}
\section{Introduction}

The ability of the simplest organisms to orchestrate basic rhythmic motor functions such as breathing, chewing, swallowing, walking or heart beating has long fascinated scientists and engineers. On the one hand, those functions have many of the attributes of autonomous clocks, suggesting that they could easily be emulated by machines. On the other hand, the adaptability and resilience of those rhythmic functions in the animal world remain to date a mystery when compared to our most advanced robots. This paradox was a central drive from the early days of cybernetics, as exemplified for instance in the homeostat of Ashby and his concept of ultrastability \citep{Ashby1952}. It remains a central question to date and was for instance heavily debated during the recent workshop on Control and Modulation of Neuronal and Motor Systems at the Mathematical Biosciences Institute. The contrast between animal and machine performance at orchestrating motor rhythms permeated much of the discussions and was equally underlined by neurophysiologists and roboticists.

This is not to say that no progress has been achieved since the days of the homeostat. Both the physiology and the mathematical modeling of rhythmic circuits is far best understood today than seventy years ago. Detailed anatomical and physiological studies of specific rhythms in specific animals have identified core neuronal mechanisms and circuit architectures that govern rhythmic functions. The  concept of central pattern generators has replaced the concept of ultrastable machines and provides a quantitative link between the neurophysiology of animal circuits and the mathematical models used to engineer rhythmic machines, see e.g. \citep{Grillner2003,Marder2007,Ijspeert2014}. The core circuit architecture of central pattern generators is an inhibitory coupling between a limited number of interneurons. Each neuron of the circuit has two distinct states of electrical activity (low and high firing rate), that ressemble the on and off state of a discrete automaton. The transition times between on and off states in each neuron is constrained by the circuit topology, which generates specific circuit rhythms via specific phase shifts between the neuronal rhythms. The core mechanims of such {\it rhythm boxes} have been extensively studied, both experimentally and computationally. The simplest such circuit  is an anti-phase rhythm between two symmetrically coupled inhibitory neurons, kwown as the half-center oscillator (HCO). Neurophysiologists have identified the specific ionic currents underlying the rebound mechanisms that generate the circuit rhythm \citep{Marder2001,Hill2003}.  Mathematical models of central pattern generators capture the phase properties of the circuits with highly simplified phase models for each neuron \citep{Ijspeert2007}. Such models have been instrumental in the development of robots that mimick animal locomotion such as \citep{Ijspeert2007,Ijspeert2014}. Collectively, those efforts have provided a comprehensive modeling framework that accounts for the biophysical principles of autonomous clocks in neuronal circuits.

How the autonomous clocks can be orchestrated with the level of adaptability and resilience found in animals remains a largely debated experimental and modeling question.
Central pattern generators, whether responsible for breathing, chewing, swallowing, walking or heart beating,  all adapt their rhythm  possibly in fractions of a second in reaction to the internal animal needs (e.g.  choking prevention) or to unexpected external contingencies (e.g.  predator escape). Growing experimental evidence suggests that such fast adaptation of  rhythmic control might happen at the cellular level, that is, without affecting synaptic strength and the circuit interconnection topology \citep{Marder2007,Harris-Warrick2011}. The role of neuromodulation has been under increasing scrutiny in the recent years. The coordination of breathing, walking, and chewing critically relies on the modulation of inhibitory interneurons excitability via monoaminergic inputs~\citep{Jordan2011,Dai2010,Liu2009,Harris-Warrick1985,Gordon2006}. More generally, neuromodulators seem to play an important role in the control of circuits via the recruitment of individual inhibitory interneurons, governing their switch from low firing activity to rhythmic bursting  \citep{McLean2008,Berkowitz2010,Zhong2011,White2011}.

Those recent experimental developments motivate the modeling question of this paper. We explore how a neuromodulatory control at the cellular level can contribute to orchestrating the rhythms of a circuit for a given connectivity. In particular, we aim at proposing a simple cellular mechanism by which the functional connectivity of a circuit is highly reconfigurable independently of its synaptic connectivity, consistently with recent experimental observations \citep{Rodriguez2013,Marder2014,Marder2015,Daur2016}. The proposed cellular mechanism is grounded in an intrinsic property of the interneurons : the presence of a slow negative conductance that can be activated by external neuromodulators. This property is provided for instance by slowly-activating calcium channels. The role of this specific intrinsic property has been extensively studied by the authors in the recent years \citep{Dethier2015,Franci2018,Drion2018}. It acts as a switch of excitability for the neuron. When the slow negative conductance is on, the excitability of the neuron is prone to slow rhythms characterized by  prolonged bursts of high frequency firing. When the slow negative conductance is off, the excitability of the neuron is only prone to the fast rhythms characteristic of individual spikes. The distinctive role of the slow negative conductance is not the generation of slow rhythms per se but rather to enable robust control and tunability of such slow rhythms. This is why it is often overlooked in computational models that concentrate on rhythm generation rather than rhythm regulation. 

In previous work, we have shown the role of the slow negative conductance in single cell control \citep{Franci2018}, in  half-center oscillator control \citep{Dethier2015}, and in the control of  large excitatory-inhibitory populations \citep{Drion2018}. Here we want to explore the role of the same slow negative conductance in orchestrating small inhibitory circuits. We use a conceptual circuit model derived after the crab somatogastric ganglion (STG) connectivity diagram (Fig. \ref{FIG:1}B, left) \citep{Gutierrez2013}. This circuit model was developed to study the interaction between two different rhythms that coexist within a same circuit. In the STG, these two rhythms are the fast pyloric rhythm, which is constantly active, and the slow gastric mill rhythm, which can be turned on and off by afferent neuromodulatory inputs and neuromodulators \citep{Marder2007}. These two rhythms are carried on by neurons that share numerous synaptic connections and many neurons switch between both rhythms or even synchronize with both rhythms at the same time \citep{Gutierrez2013,Bucher2006,Dickinson1990,Meyrand1991,Weimann1994}

Our results show that a slow negative conductance endows each cell of the circuit with a switching mechanism  that governs their individual participation in the circuit rhythm and that is externally controlled by neuromodulators.  We constrast the controllability properties of the circuit in the presence and in the absence of the cellular switch. A weakly interconnected circuit with switchable neurons is shown to be robust, adaptable and reconfigurable. The synaptic connectivity constrains the phase relationships between the neurons, as in the classical models of central pattern generators. But the participation of each neuron in the orchestra is controlled externally and allows for a continuous and robust modulation of the circuit rhythms. The distinctive property of the slow negative conductance is that it only controls the switch but does not constrain the rhythm. In this manner, our model suggests a way to reconcile the clock properties of an autonomous rhythm box with the adaptation and resilience properties of rhythmic motor functions. We propose that the cellular switching mechanism can be easily incorporated in abstract computational models of central pattern generators and that  it is important to account for this distinctive feature when studying the modulation and controllability properties of a rhythmic circuit.

\section{Results}
\subsection{A source of cellular slow negative conductance is critical for the coexistence of switchable rhythms within a same circuit}
\begin{figure}
\centering
\includegraphics[width=0.9\textwidth]{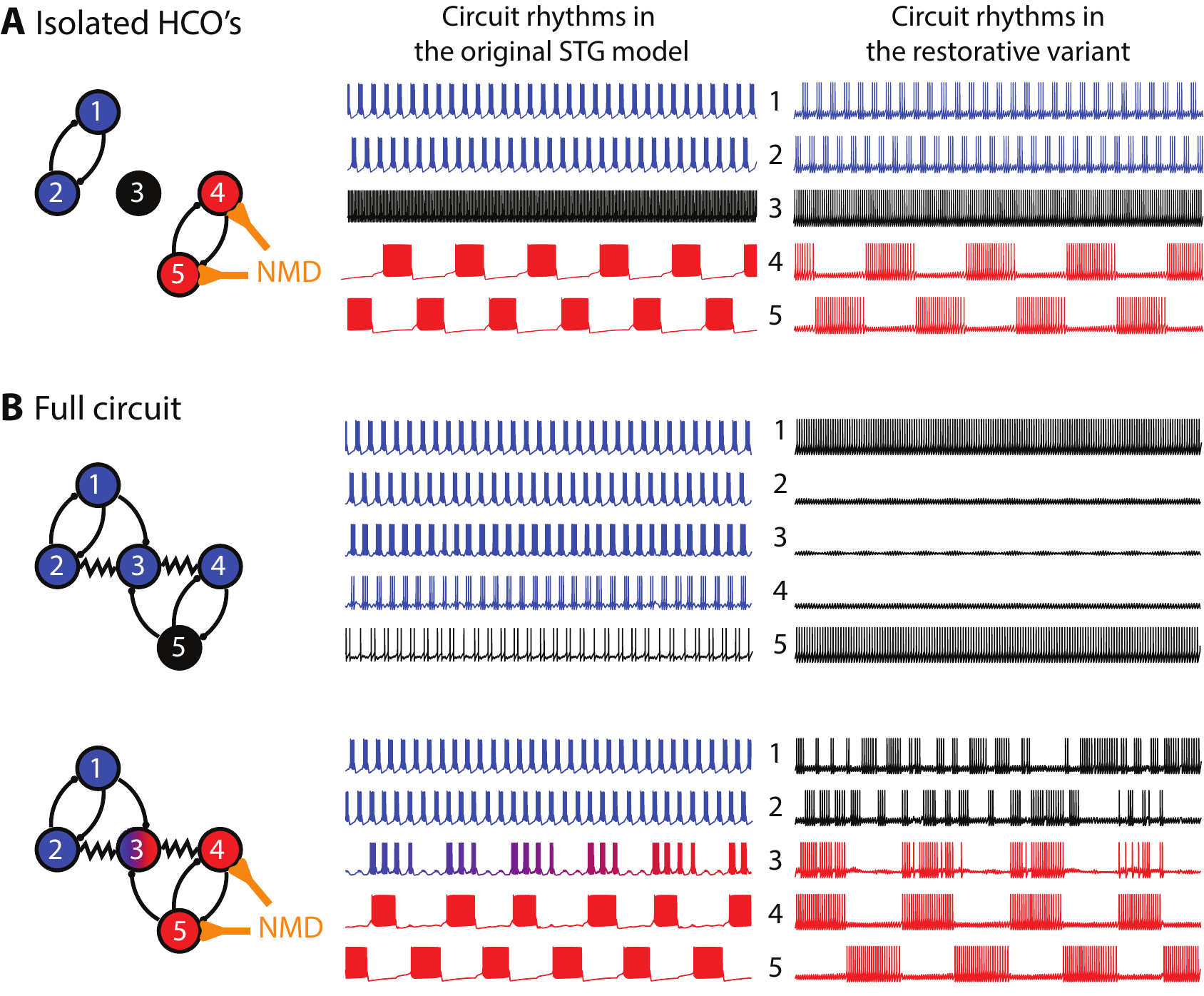}
\caption{\textbf{Only the circuits  that include a cellular slow negative conductance have robust rhythms modulated by connectivity changes and external inputs.} \textbf{left}, circuit connectivity diagram for (A) isolated half-center oscillators (HCO's)  and (B)  full connectivity  with or without modulatory input. Filled circles represent neurons, which are numbered from 1 to 5. Black curves with rounded heads represent inhibitory connections. Tick orange curves represent active neuromodulatory pathways targeting neurons 4 and 5. Resistor symbols represent electrical synaptic connections. Neurons that are involved in a fast rhythm are highlighted in blue, neurons that are involved in a slow rhythm are highlighted in red, and neurons that are not involved in a circuit rhythm are highlighted in black.  \textbf{center and right} Membrane potential variations over time for neurons 1 to 5 (from top to bottom). Circuits in the left column include a slow negative conductance and are weakly connected. Circuits on the right do not include a slow negative conductance and require strong connectivity.  A comparison between center and right traces shows that only circuits that include a cellular slow negative conductance allow for  coexisting fast and slow rhythms.}\label{FIG:1}
\end{figure}

Throughout this paper, we analyse a specific circuit model composed of five interconnected neurons. Neurons~1,2 and neurons~4,5 are respectively connected through mutually inhibitory synapses (Fig.~\ref{FIG:1}A). Each pair of neurons  generates the  anti-phase rhythmic activity of a half center oscillator (HCO). Following the approach of~\citep{Gutierrez2013}, the HCO composed of neurons~1,2 is designed to generate a fast and continuously active, pyloric-like rhythm (blue traces), whereas the HCO composed of neurons~4,5 is designed to generate a slow, gastric mill-like rhythm that is only active under the action of specific neuromodulators (NMD) (red traces). Neuron 3 is a hub neuron, connected to both HCOs through electrical and chemical synapses as in Fig.~\ref{FIG:1}B, left.

The dynamics of each neuron were modeled using two variants of the STG neuron conductance-based model described in \citep{Liu1998}. This model is composed of seven voltage-gated currents, two of which are sources of a slow negative conductance (the two slowly-activating calcium currents $I_{CaT}$ and $I_{CaS}$). In the first variant (called the \emph{original model}), the kinetics originally described in \citep{Liu1998} were used, making the modulation of the cellular slow negative conductance possible through the modulation of calcium channel density \citep{Franci2018,Drion2015}. In the second variant (called the \emph{restorative variant}), the activation of both types of calcium channels were made ten times faster. This modification did not affect the steady-state properties of the model, both models having identical IV curves for any given parameter set. The slow negative conductance is however lost in the restorative variant because the two calcium conductances are now sources of fast rather than slow negative conductance. The meaning of {\it fast} and {\it slow} is relative to the kinetics of sodium activation, here considered as {\it fast}. In the original version of the model, the kinetics of calcium activation is ten times slower than the kinetics of sodium activation, whereas the two distinct time scales are merged in the restorative variant. The reader is referred to \citep{Franci2018} for a detailed analysis of the dynamical consequences of this difference at the single cell level.

First, we analyzed the ability to generate fast and slow rhythms in isolated, symmetrical and homogenous HCOs (i.e. with no connection to the hub neuron) via different mechanisms (Fig.~\ref{FIG:1}A). A parameter space exploration shows that many different parameter combinations can lead to HCO rhythms. These rhythms can however be grouped into two categories according to wether their origin mainly relied on cellular properties or on (strong) circuit interconnection.

On the one hand, fast and slow HCO's can be generated by exploiting the cellular sources of slow negative conductance: the presence of a slow negative conductance at the cellular level make the cells prone to engage in a circuit rhythmic activity, even for very weak synaptic connections (Fig.~\ref{FIG:1}A, center). This {\it cellular mechanism} is accessible to all neurons or models that possess at least one source of slow negative conductance (i.e. one slowly activating inward current or one slowly inactivating outward current). Yet it has not received a lot of attention until very recently, even in the context of HCO rebound mechanisms~\citep{Dethier2015}.

On the other hand, HCO's can be generated in the absence of any cellular slow negative conductance using strong synaptic connections. The synaptic connection strength in Fig.~\ref{FIG:1}A right is ten times the synaptic connection strength in Fig.~\ref{FIG:1}A center. This {\it circuit mechanism} is well known and has been extensively studied in the past \citep{Marder2001,Hill2003}. It is the only mechanism that can be used to build HCOs made of neuron models lacking any source of slow negative conductance, which includes any detailed conductance-based model with fast/instantaneous calcium activation \citep{Terman2002,Rubin2004,Butera1999a,Golomb1997} and simple spiking neuron models such as the FitzHugh-Nagumo model \citep{FitzHugh1961} and integrate-and-fire models \citep{Gerstner2014}.

Secondly, we analyzed the robustness and possible coexistence of the two HCO rhythms in the five neuron model (Fig.~\ref{FIG:1}B) when HCO rhythms relies on either the cellular or circuit mechanism. As described above, the only difference between the two configurations is the calcium current activation time constant and the synaptic connection strength. This ensures that the cells only differed by the presence or absence of sources of a slow negative conductance, all other properties such as membrane input resistance or neuron IV curve being identical. In the isolated HCOs (Fig. \ref{FIG:1}A), both the cellular or circuit mechanisms can produce very similar circuit rhythms in terms, for instance, of period and duty cycle. In both cases, the slow rhythm can be turned ON by the presence of a neuromodulator (NMD) that increases calcium channel densities. However, the two mechanisms show very different robustness properties when involved in the larger circuit (Fig. \ref{FIG:1}B).

The cellular mechanism produces HCO rhythms that are robust to changes in circuit connectivity and  modulatory state of the slow HCO. With the slow rhythm turned OFF, the fast HCO rhythm is barely affected by its interconnection with the other neurons in the circuit, and even propagates throughout the circuit by engaging both the hub neuron and one neuron of the slow HCO in the fast rhythm (Fig. \ref{FIG:1}B, top center). The non-rhythmic neurons are entrained by the robust HCO rhythm. Modulating the slow HCO turns the slow rhythm ON without disrupting the fast rhythm, and both rhythms coexist at the level of the hub neuron (Fig. \ref{FIG:1}B, bottom center). Such robust switches in circuit activity are reminiscent of what is observed in the STG \citep{Marder2007} and other central pattern generators \citep{Harris-Warrick2011}.

The circuit mechanism, on the other hand, produces HCO rhythms that are strongly affected by the full, 5 neuron circuit interconnection. With the slow rhythm turned OFF, the rhythm generated by the fast HCO does not survive its interconnection with the other neurons in the circuit even for relatively weak connectivity (in the example shown in Fig.~\ref{FIG:1}B, top right, the connections to the hub neuron are 10 times weaker than the connections within the HCO's, whereas they are of similar amplitude in the previous case). In this case, the non-rhythmic neurons disrupt the HCO activity instead of being entrained by it. Maintaining the HCO rhythm requires to isolate the HCO sub-circuit by using extremely weak connections to the hub neuron, and there is no propagation of the rhythm throughout the circuit. Modulating the slow HCO turns the slow rhythm ON but results in some erratic behaviors in the fast neurons (Fig. \ref{FIG:1}B, bottom right). Fast and slow rhythms do not coexist robustly within the same circuit.

These results indicate that, although both cellular and circuit mechanisms can generate rhythms in simple circuit configurations, the cellular mechanism, which relies on the modulation of cellular slow negative conductances, is required for the robustness and coexistence of rhythms in larger connectivity diagrams. This highlights the important role played by slow regenerative channels for the generation and modulation of circuit rhythms in central pattern generators.

\subsection{In spite of a fixed connectivity, cellular control of the slow negative conductances orchestrates the circuit rhythms}
The previous section showed that circuit rhythms created by symmetrical HCO's could robustly coexist within a more complex structural connectome if they relied on the presence of a slow negative conductance at the cellular level. Single rhythms could however still be generated in the full structural connectome using strong connectivity between restorative neurons in this specific set-up. In this section, we further explore the robustness and rhythmic capabilities of both mechanisms in circuits without any predefined structure in neuronal dynamics and synaptic connection strength. To this end, we simulated circuits with random voltage-gated current densities and random synaptic connection strengths (all were randomly picked in a range $[\bar{g}_{c} - 25\%, \bar{g}_{c} + 25\%]$ following a uniform distribution, where $\bar{g}_{c}$ is a central value that was chosen for each current following the results presented in Fig. \ref{FIG:1}). Furthermore, for each defined cellular and circuit configuration, several neuromodulatory states were simulated. 

In the absence of a slow negative conductance at the cellular level, breaking network symmetry and increasing cellular heterogeneity results in the incapacity to generate any circuit rhythm for most randomly picked parameter sets and neuromodulatory states, even with strong network connectivity. We do not show any quantitative result here, but identical observations are reported in variable circuits and heterogeneous networks in \citep{Dethier2015,Drion2018}.

\begin{figure}
\centering
\includegraphics[width=0.9\textwidth]{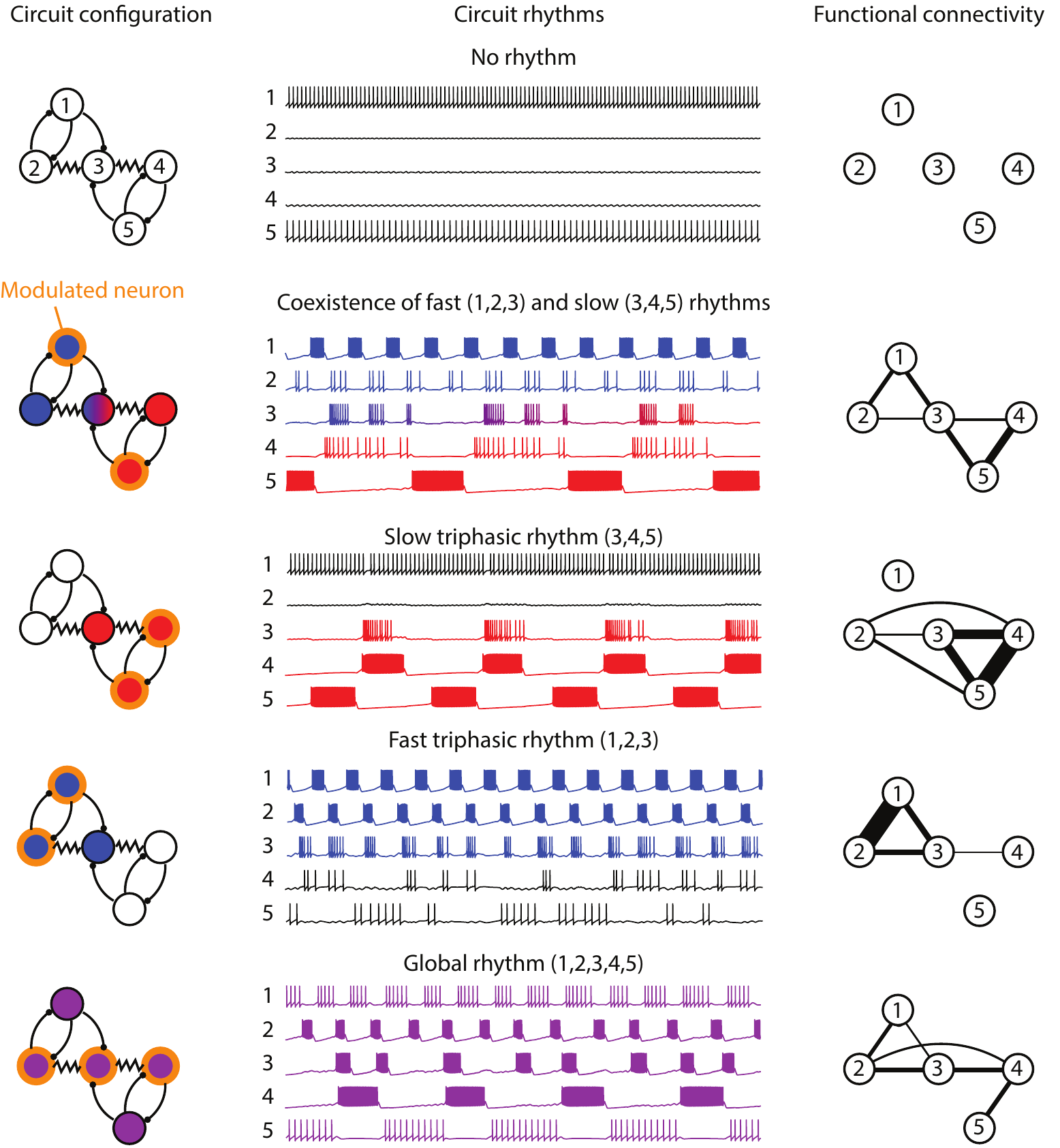}
\caption{\textbf{Modulating the slow negative conductances of specific neurons selects different circuit rhythms and functional connectivities exist in spite of a fixed synaptic connectivity.} Left, circuit connectivity diagrams. Filled circles represent neurons, which are numbered from 1 to 5. Black curves with rounded heads represent inhibitory connections. Resistor symbols represent electrical synaptic connections. Neurons with tick orange edges are subject to neuromodulatory inputs that increase the voltage-gated calcium channel density. Center, membrane potential variations over time for neurons 1 to 5 (from top to bottom) in the different modulatory configurations. Synaptic connections are identical in all cases.
Neurons are colored in blue when they participate to the fast rhythm,  in red when they participate in the slow rhythm, in purple when they participate in a global rhythm, and in black when they do not participate to the circuit rhythm.  Right, functional connectome in the different modulatory configurations. Line thickness is proportional to the maximum of the cross covariance function between voltages after low-pass filtering. No line was drawn if this maximum was lower than $10$.}\label{FIG:2}
\end{figure}

On the other hand, breaking network symmetry and increasing cellular heterogeneity in the presence of slow negative conductance at the cellular level revealed the ability of a fixed structural connectome to generate rich and diverse circuit rhythms or combinations of circuit rhythms. An illustrative example of this ability is provided in Fig. \ref{FIG:2}. The figure shows the activity of one of the circuits obtained by random variations of the model parameters in 5 different neuromodulatory states. All parameters are identical in all states except for neuron calcium channel densities, which are under the control of external neuromodulators, hence the structural connectome is identical in each state. In the absence of any neuromodulartory input, the slow negative conductance is shut down in all neurons and the circuit exhibit no rhythmic activity (Fig. \ref{FIG:2}, top). However, activating neuromodulatory pathways turn ON circuit rhythms of variable types depending on the set of neurons that are targeted by neuromodulators. In the provided example, the fixed structural connectome could exhibit either two coexisting fast and slow rhythms, a slow triphasic rhythm, a fast triphasic rhythm, and a global network rhythm for 4 different neuromodulatory states (Fig. \ref{FIG:2}, from second to bottom). 

We performed a {\it functional connectivity} experiment on the toy circuit of Fig.~\ref{FIG:2} left. We low-pass filtered membrane potential traces and measured the cross covariance between nodes in the circuit. This provides a black-box representation of the effective connectivity, or ``functional connectome''. The results are reported in Fig.~\ref{FIG:2} right, with thicker lines corresponding to higher cross covariance. The rhythmic state of the circuit is trustfully reflected in the measured functional connectome. For instance, in the fast (resp. slow) triphasic rhythm state the functional connectivity highlights the underlying three neuron core. Both the coexistence of the fast and slow rhythms and the global rhythm lead to a connected functional connectivity graph. The quantitative difference is that the main connecting core is the neuronal triple $1-3-5$ in the case of coexistence of fast and slow rhythms and the triple $2-3-4$ in the case of the global rhythm.

The simple experiment in Fig.~\ref{FIG:2} highlights a fundamental fact. A fixed synaptic interconnection topology, with fixed synaptic connection strengths is able to support a rich dictionary of possible functional connectivity motifs, provided that tunable slow negative conductances are available at the cellular level.

Figure \ref{FIG:2} only shows a few examples of the many network activities that can be exhibited by the circuit with fixed synaptic connections. But in a 5-cells circuit and considering that each neuron can switch between two states (slow negative conductance OFF vs slow negative conductance ON), the circuit can potentially reach $2^5 = 32$ different neuromodulatory states, each potentially resulting in its own specific circuit rhythm or combination of circuit rhythms. Considering the more physiological assumption that cellular slow negative conductances can be modulated in a continuous way, depending on neuromodulator concentration for instance, further enriches the rhythmic capabilities of the circuit. Furthermore, rhythms that rely on the cellular mechanism being robust to changes in intrinsic parameters that do not affect the slow negative conductance, each neuromodulatory state can specifically be regulated by targeting voltage-gated currents that are not the principal sources of a slow negative conductance. Finally, modulation of synaptic connection (not considered here) adds a further layer of rhythm regulation that is compatible with the slow negative conductance switch.

\subsection{Simple modeling of circuit rhythm modulation}
Detailed neuron conductance-based models have the advantage of closely relating to physiology. They however can rapidly grow in dimension as further details about the neuron under study are taken into consideration. They often contain tens to hundreds of cell specific parameters, most of which are unknown, and model behavior can be highly sensitive to changes in many of them \citep{Goldman2001,Prinz2003}, making the parametrization of a conductance-based model an arduous case-by-case task. Results obtained using one specific conductance-based model are therefore often highly dependent on model specifics, hence difficult to generalize. 

So far, we have shown that a change in one single parameter in a specific neuron conductance-based model that contains about hundred different parameters disrupts the whole 
control of the rhythmic circuit. This  parameter controls the activation kinetics of  the two calcium currents of the model, which are the only two  sources of a slow negative conductance in the model. We concluded that switching slow negative conductance ON and OFF at the cellular level was key to the robustness of rhythms at the circuit level. In order to further establish the generality of these results, we reproduced the previous experiments using a simple, hybrid model that captures the core dynamics of arbitrary conductance-based models. The simple model reads
\begin{eqnarray}
\dot{V} &=& V^2 - x_s^2 + bVx_s - g_sx_s - g_ux_u + I_{app} \\ 
\dot{x_s} &=& \varepsilon_s(a_sV - x_s) \\
\dot{x_u} &=& \varepsilon_u(a_uV - x_u)
\end{eqnarray}
where $V$ is the membrane potential, $x_s$ a slow recovery variable and $x_u$ an ultraslow recovery variable ($\varepsilon_s >> \varepsilon_u$). $V$ and $x_s$ are reset to $V_{reset}$ and $x_{s,reset}$, respectively, each time $V$ crosses a threshold value $V_{th}$. In Equation (1), the term $V^2 - x_s^2 + bVx_s$ is based on the local normal form of a transcritical bifurcation, which has been shown to organize the transition between \emph{restorative excitability}, characterized by the absence of a slow negative conductance, and \emph{regenerative excitability}, characterized by the presence of a slow negative conductance \citep{Drion2012,Franci2013}. In our model, this transition is controlled by the parameter $g_s$, which controls the amplitude and sign of the slow conductance. This parameter captures the role played by the slowly activating calcium currents in the conductance-based model used above. 

\begin{figure}
\centering
 \includegraphics[width=0.75\textwidth]{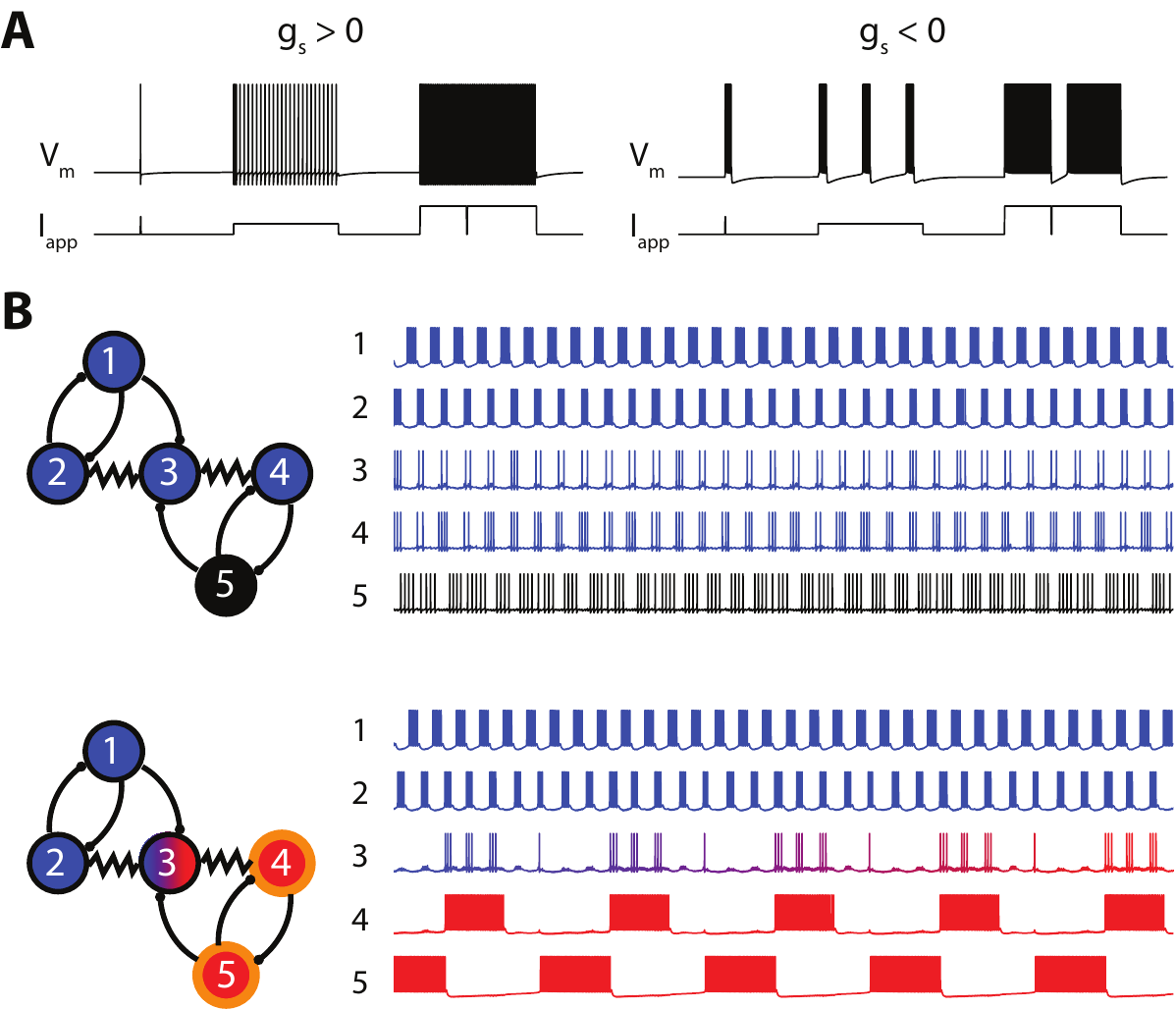}
\caption{{\bf Coexistence of rhythms in a simplified computational model that captures the role of the slow negative conductances via the single conductance parameter $g_s$.} \textbf{A,} Responses of the hybrid neuron model ($V_m$, top) to pulses of depolarizing and hyperpolarizing current ($I_{app}$, bottom) when the slow conductance around resting potential is positive ($g_{s}>0$, left) or negative ($g_{s}<0$, left). All other parameters are identical, expect for a steady current that decouples the value of $g_s$ and model excitation state. \textbf{B, left,} circuit connectivity diagrams. Filled circles represent neurons, which are numbered from 1 to 5. Black curves with rounded heads represent inhibitory connections. Resistor symbols represent electrical coupling. Tick orange curves represent active neuromodulatory pathways targeting neurons 4 and 5. \textbf{B, right,} membrane potential variations over time of each neuron in the absence (top) or presence (bottom) of a neuromodulatory input to neurons 4 and 5. Neurons are colored in blue when they participate in the fast rhythm, in red when they participate in the slow rhythm, and in black when they do not participate  in the circuit rhythm.}\label{FIG:3}
\end{figure}

The effect of $g_s$ on model excitability is illustrated in Fig. \ref{FIG:3}A. The figure shows model response to short and long-lasting pulses of depolarizing and hyperpolarizing current for a positive or negative value of $g_s$ (i.e. in the absence or presence of a slow negative conductance). All other parameters are strictly identical, expect for a steady current that decouples the value of $g_s$ and model excitation state (see Methods). For positive $g_s$, the model shows the signatures of \emph{restorative excitability}: a short pulse of depolarizing current induces a single spike, a long lasting pulse induces sustained spiking whose frequency depends on the amplitude of the depolarizing current (Fig. \ref{FIG:3}A, left). Such responses are reminiscent of already available simple neuron models. For negative $g_s$, the model response strongly differs and shows the signatures of \emph{regenerative excitability}: a short pulse of depolarizing current induces a burst of spikes, a long lasting pulse of moderate amplitude induces bursting, and a long lasting pulse of large amplitude induces sustained spiking that can be temporarily interrupted by a short hyperpolarizing pulse (Fig. \ref{FIG:3}A, right).

Figure \ref{FIG:3}B shows the circuit rhythm generated by the interconnection of the simple models in the two modulatory states considered in Fig. \ref{FIG:1}B. Both the connectivity diagram and synapse models are identical to the ones used in the previous sections. In the absence of a slow negative conductance in neurons 4 and 5 (i.e. neuromodulatory inputs are shut down), the fast HCO rhythm generated by neurons 1 and 2 robustly spreads throurough the circuit up to neurons 3 and 4 (Fig. \ref{FIG:3}B, top). Turning ON the slow negative conductance in neurons 4 and 5 induces the generation of a slow HCO rhythm that coexists with the fast rhythm, and neuron 3 is involved in both the fast and the slow rhythms. 

These results show that, although the simple model lacks most of the specific details of the conductance-based, it is sufficient to capture the cellular mechanism necessary to the  robust control of the circuit rhythms. The significance of such results is twofold. On the one hand, they demonstrate the critical role played by the cellular slow negative conductance for the generation and control of robust circuit rhythms. On the other hand, they show that robust circuit rhythmic activity can be studied using simple neuron models provided that these models have the ability to control the absence or presence of a slow negative conductance at the cellular level. Several such models have been recently developed and studied mathematically \citep{Drion2012,Franci2012,Franci2014,VanPottelbergh2017}.

\section{Discussion}
\subsection{Decoupling the rhythm selection from the rhythm regulation}
We have highlighted the role of a   slow negative  cellular conductance  in the robust control of circuit rhythms. This property endows each neuron with two distinct modes of excitability : the neuron is burst excitable when the slow negative conductance is on and spike excitable when the slow negative conductance is off. This switching mechanism is simple. It exists in any neuron that includes slowly-activating calcium currents and its external control only requires a neuromodulator that controls channel density. Recent results suggest experimental evidence of neuromodulators that modulate the slow negative conductance of the neuron and participate to the control of the circuit rhythm \citep{McLean2008,Berkowitz2010,Zhong2011,White2011}. The cellular switch does not tune the rhythm. It only controls the participation of a given node in the network rhythm. 

The cellular mechanism endows the circuit with a selection mechanism: many functional connectomes can be selected from a given structural connectome, that is, a given configuration of synaptic and electrical couplings.  Neuromodulators can selectively control the participation of each node in the functional connectome, considerably enriching the repertoire of possible circuit rhythms.  Theoretically, the number of possible circuit states is combinatorial in the number of controlled nodes.

The selection of a functional connectome is largely decoupled from the tuning of the circuit rhythm in that particular configuration. The circuit rhythm results from specific transition times between high and low firing activity of the neurons and the functional connectome only constrains the phase properties of those transition times. The precise tuning of the transition times involves both the intrinsic properties of the neurons and the synaptic and electrical couplings. In that sense, the rhythm regulation is largely decoupled from the rhythm selection.

This decoupling is required by the different time scales involved in circuit regulation. The behavioral time scale of an animal requires control actions within hundreds of milliseconds to seconds. Such rapid modulation is not compatible with synaptic plasticity. But it is compatible with the action of neuromodulators that select different functional connectomes. There is increasing experimental evidence  that neuromodulation plays a major role in shaping the functional connectome \citep{Marder2012,Bargmann2013}. This conclusion probably extends to many more spatial scales, from small circuits to whole brain activity \citep{Haider2009,Mennes2012,Ekman2012,Francis2018,Honey2007}. 

 The tuning of circuit rhythms occurs over a broad range of time scales that include for instance the slower time scale of synaptic plasticity. 
While the tuning properties of the circuit rhythms have not been studied in detail in the present paper, it is rather intuitive that a discrete selection mechanism considerably enriches the tunability of a circuit. Our previous work \citep{Dethier2015} showed how the slow negative conductance of neurons contributes to robustness and tunability of a half-center oscillator, which is directly relevant to the circuits studied in the present paper. Our recent work \citep{Drion2018}  shows similar conclusions in much larger populations of neurons that switch between an active and oscillatory state.

\subsection{A cellular property that matters at the circuit level}
The slow negative conductance emphasized in this paper is not 
the property of a particular ionic current. It is a property of the total current-voltage relationship of the neuron and it can be regulated by the expression of many different channels. It has a specific signature in a voltage-clamp experiment (see for instance the discussion in \citep{Franci2018}) and can be easily assessed in a detailed conductance-based model (see for instance \citep{Franci2013}). This property is also distinct from the rebound mechanism property necessary for the antiphase rhythm of a half-center oscillator \citep{Dethier2015}. It is in fact a property that seems to have received little attention until recently \citep{Drion2012}.

The question of which cellular details must be included in the computational model of a circuit is often a matter of debate: detailed cellular models facilitate the biophysical interpretation of the model properties but detailed cellular models are impractical in circuit studies because they result in large dimensional models with many parameters to tune rather arbitrarily. Circuit models that use a simplified neuronal model are thus preferred but they raise the question of which details of the full model must be retained. 

Popular simplified neuronal models include the Hodgkin-Huxley model, planar reductions of Hodgkin-Huxley model (e.g. Fitzhugh-Nagumo model), or linear and quadratic-integrate and fire models. None of those models possess the cellular property emphasized in this paper. We refer the interested reader to \citep{Drion2012,VanPottelbergh2017} for a detailed analysis of why those models do not include a slow negative conductance and how to modify them to account for this specific property.

We have illustrated the control properties of the STG circuit by using a detailed conductance based model for each neuron, but we have also shown that the same control mechanism can be reproduced in a reduced model that uses highly simplified equations for each neuron. This simplification underlines that the proposed mechanism does not result from the  biological details of  neuronal models but only from the slow negative conductance property of the total current-voltage relationship of each neuron. It is therefore conceivable to include this property in simple models of artificial central pattern generators that could inform the design and control principles of rhythmic machines. Such models could for instance expand the control principles grounded in phase models of central pattern generators in the design of artificial locomotion \citep{Ijspeert2007}.

\section{Methods}
\subsection{Neuron conductance-based model}
All simulations and analyses were performed using the Julia programming language. The Julia code is freely available at \\http://www.montefiore.ulg.ac.be/$\sim
$guilldrion/Files/DFS2018-code.zip.

Figures \ref{FIG:1} and \ref{FIG:2} were generated using the STG model described in \citep{Liu1998}. The model follows the voltage equation
$$
C\dot{V} = -I_{ion} + I_{app}
$$
where $V$ is neuron membrane potential, $C$ is the membrane capacitance, $I_{ion}$ represents ionic currents and $I_{app}$ represents an externally applied current. Ionic currents are composed of a leak current $I_{leak}$, a transient sodium current $I_{Na}$, a T-type calcium current $I_{Ca,T}$, a S-type calcium current $I_{Ca,S}$, a delayed rectifier potassium current $I_{K,DR}$, a transient potassium current $I_{A}$, a calcium activated potassium current $I_{K,Ca}$ and a hyperpolarization-activated cation current $I_{H}$. Fixed parameters used in the simulations were as follows: $C=1\,\mu F\cdot cm^{-2}$, $V_{Na}=50\,mV$, $V_K=-80\,mV$, $V_{Ca}=80\,mV$, $V_{leak}=-50\,mV$, and $\bar g_{leak}=0.01\,mS\,cm^{-2}$.

In Figure \ref{FIG:1}, variable parameters were as follows. Neurons 1 and 2: , $\bar g_{Na}=600\,mS\,cm^{-2}$, $\bar g_{Ca,T}=3\,mS\,cm^{-2}$, $\bar g_{Ca,S}=8\,mS\,cm^{-2}$, $\bar g_{A}=50\,mS\,cm^{-2}$, $\bar g_{K,DR}=90\,mS\,cm^{-2}$, $\bar g_{K,Ca}=60\,mS\,cm^{-2}$. Neuron 3: $\bar g_{Na}=600\,mS\,cm^{-2}$, $\bar g_{Ca,T}=3\,mS\,cm^{-2}$, $\bar g_{Ca,S}=2\,mS\,cm^{-2}$, $\bar g_{A}=50\,mS\,cm^{-2}$, $\bar g_{K,DR}=90\,mS\,cm^{-2}$, $\bar g_{K,Ca}=30\,mS\,cm^{-2}$. Neurons 4 and 5: $\bar g_{Na}=600\,mS\,cm^{-2}$, $\bar g_{A}=50\,mS\,cm^{-2}$, $\bar g_{K,DR}=90\,mS\,cm^{-2}$, $\bar g_{K,Ca}=60\,mS\,cm^{-2}$, and $\bar g_{Ca,T}=1\,mS\,cm^{-2}$, $\bar g_{Ca,S}=1\,mS\,cm^{-2}$ (unmodulated state) or $\bar g_{Ca,T}=3\,mS\,cm^{-2}$, $\bar g_{Ca,S}=8\,mS\,cm^{-2}$ (modulated state). In addition, a parameter $\tau_{mK,Ca}$ multiplying the calcium-activated potassium current time-constant was added and set to $1$ for neurons 1, 2 and 3 (fast rhythm) and set to $20$ for neurons 4 and 5 (slow rhythm). 

In Figure \ref{FIG:2}, maximal conductances were randomly picked in a range $[\bar{g}_{c} - 25\%, \bar{g}_{c} + 25\%]$ following a uniform distribution, where $\bar{g}_{c}$ is a central value that was chosen for each current. These central values were $\bar g_{Na,c}=600\,mS\,cm^{-2}$, $\bar g_{A,c}=50\,mS\,cm^{-2}$, $\bar g_{K,DR,c}=90\,mS\,cm^{-2}$, $\bar g_{K,Ca,c}=60\,mS\,cm^{-2}$. For the unmodulated state: $\bar g_{Ca,T,c}=1\,mS\,cm^{-2}$, $\bar g_{Ca,S,c}=3\,mS\,cm^{-2}$. For the modulated state: $\bar g_{Ca,T,c}=3\,mS\,cm^{-2}$, $\bar g_{Ca,S,c}=8\,mS\,cm^{-2}$. $\tau_{mK,Ca}$ was randomly picked in a range [5.5,24.5]. 

Finally, a parameter $\tau_{mCa}$ multiplying both calcium current time-constants was added and set to $1$ for the \emph{original model} and to $0.1$ to create the \emph{restorative variant}.

\subsection{Neuron hybrid model}
The neuron hybrid model used in Figure \ref{FIG:3} is based on the normal form of the transcritical bifurcation \citep{Drion2012,Franci2013}. It reads
\begin{eqnarray}
\dot{V} &=& V^2 - x_s^2 + bVx_s - g_sx_s - g_ux_u + I_{ss} + I_{app} \nonumber \\ 
\dot{x_s} &=& \varepsilon_s(a_sV - x_s) \nonumber \\
\dot{x_u} &=& \varepsilon_u(a_uV - x_u)\nonumber 
\end{eqnarray}
where $V$ is the membrane potential, $x_s$ a slow recovery variable, $x_u$ an ultraslow recovery variable ($\varepsilon_s >> \varepsilon_u$) and $I_{app}$ an externally applied current. $I_{ss} = (-(-2V_{ss}(1-a_s^2+a_sb)-a_s(g_s+g_u))^2+a_s^2(g_s+g_u)^2)/(4(1-a_s^2+a_sb))$ is a steady current that decouples the value of $g_s$ and model excitation state at $V_{ss}$ (set to -2). To make the value of the membrane potential similar to the one of the conductance-based model, the membrane potential $V$ was shifted by $-70\,mV$. This shift is only necessary to ensure similar synaptic current activation with both models. Finally,

The reset rule reads
\begin{eqnarray}
\mbox{if } V > V_{th} \quad \mbox{ then } \quad & V & \leftarrow V_{reset}\nonumber \\ 
& x_s & \leftarrow x_{s,reset}\nonumber \\
& x_u & \leftarrow x_{u} + \Delta x_{u}.\nonumber 
\end{eqnarray}

The non tunable parameters were set as follows: $b = -2$, $a_s = a_u = 0.1$, $V_{reset} = 40$, $x_{s,reset} = 30$ and $\Delta x_{u} = 20$. The tunable parameters were neuron dependent. Neurons 1 and 2: $g_s = -30$, $g_u = 2$, $I_{app} = 40$, $\varepsilon_s = 1$, $\varepsilon_u = 0.1$. Neuron 3: $g_s = 0$, $g_u = 1$, $I_{app} = 0$, $\varepsilon_s = 1$, $\varepsilon_u = 0.1$. Neurons 4 and 5: $g_s = 0$ (unmodulated state) or $g_s = -30$ (modulated state), $g_u = 0.2$, $I_{app} = 60$, $\varepsilon_s = 1$, $\varepsilon_u = 0.01$.

\subsection{Synaptic connections}
Neurons were connected through chemical inhibitory synapses and electrical synapses. Synaptic currents were added to the voltage equation following 
$$
C\dot{V} = -I_{ion} + I_{app} -I_{syn} -I_{el} 
$$
in case of the conductance-based model, and
$$
\dot{V} = V^2 - x_s^2 + bVx_s - g_sx_s - g_ux_u + I_{ss} + I_{app} -I_{syn} -I_{el}
$$
in case of the hybrid model, where $I_{syn}$ represents the chemical synaptic current and $I_{el}$ the electrical synaptic current. The electrical synaptic current was modeled following the equation $I_{el}=\bar{g}_{el}(V_{post} - V_{pre})$ where $\bar{g}_{el}$ is the maximal conductance of the electrical synapse, $V_{post}$ is the membrane potential of the postsynaptic neuron and $V_{pre}$ is the membrane potential of the presynaptic neuron. Electrical synapses were considered to transmit in both directions. The chemical synaptic current was modeled following the equation $I_{syn}=\bar{g}_{syn}s(V_{post} - V_{syn})$ where $\bar{g}_{syn}$ is the maximal conductance of the chemical synapse, $V_{post}$ is the membrane potential of the postsynaptic neuron and $V_{syn}$ is the synaptic reversal potential (set to $-75 mV$ in all simulations). $s$ is the synaptic activation variable that depends on the membrane potential of the presynaptic neuron following the equation
$$
\tau_s \dot{s} = s_\infty(V_{pre})-s
$$
where $\tau_s$ is the synaptic activation time-constant, which was set to $10\,ms$ when using the conductance-based models and to $1\,ms$ when using the hybrid model (although the dimension has little meaning in the latter case). $s_\infty(V_{pre})$ is the steady-state activation curve defined by
$$
s_{\infty} = \left\{ \begin{array}{ll}
0 &\quad \mbox{if $V_{pre} < V_{th,syn}$};\\
\tanh((V_{pre}-V_{th,syn})/V_{slope}) &\quad \mbox{if $V_{th,syn} \geq V_{th,syn}$}.\end{array} \right.
$$
where $V_{th,syn}$ was set to $-50\,mV$ and $V_{slope}$ to $10\,mV$. 

In Figure \ref{FIG:1}A, center, the following synaptic maximal conductances were used ($\bar{g}_{syn}^{i,j}$ represents the synaptic connection from neuron $i$ to neuron $j$): $\bar{g}_{syn}^{1,2} = \bar{g}_{syn}^{2,1} = \bar{g}_{syn}^{4,5} = \bar{g}_{syn}^{5,4} = 0.2\,mS\,cm^{-2}$, $\bar{g}_{syn}^{1,3} = \bar{g}_{syn}^{5,3} = \bar{g}_{el}^{2,3} = \bar{g}_{el}^{3,2} = 0\,mS\,cm^{-2}$. In Figure \ref{FIG:1}A, right, the non-zero maximal synaptic conductances were ten times stronger: $\bar{g}_{syn}^{1,2} = \bar{g}_{syn}^{2,1} = \bar{g}_{syn}^{4,5} = \bar{g}_{syn}^{5,4} = 2\,mS\,cm^{-2}$. In Figure \ref{FIG:1}B, synaptic maximal conductances were as in Figure \ref{FIG:1}A except for $\bar{g}_{syn}^{1,3} = \bar{g}_{syn}^{5,3} = 0.2\,mS\,cm^{-2}$ and $\bar{g}_{el}^{2,3} = \bar{g}_{el}^{3,2} = 0.05\,mS\,cm^{-2}$. 

In Figure \ref{FIG:2}, synaptic maximal conductances were randomly picked in a range $[\bar{g}_{c} - 25\%, \bar{g}_{c} + 25\%]$ following a uniform distribution, where $\bar{g}_{c}$ is a central value chosen as follows: $\bar{g}_{syn,c}^{1,2} = \bar{g}_{syn,c}^{2,1} = \bar{g}_{syn,c}^{4,5} = \bar{g}_{syn,c}^{5,4} = \bar{g}_{syn,c}^{1,3} = \bar{g}_{syn,c}^{5,3} =0.01\,mS\,cm^{-2}$ and $\bar{g}_{el,c}^{2,3} = \bar{g}_{el,c}^{3,2} = 0.02\,mS\,cm^{-2}$. 

In Figure \ref{FIG:3}, the following synaptic maximal conductances were used: $\bar{g}_{syn}^{1,2} = \bar{g}_{syn}^{2,1} = \bar{g}_{syn}^{4,5} = \bar{g}_{syn}^{5,4} = \bar{g}_{syn}^{1,3} = \bar{g}_{syn}^{5,3} = 15$ and $\bar{g}_{el}^{2,3} = \bar{g}_{el}^{3,2} = 3$. These values are dimensionless. 

\subsection{Functional connectome}

The functional connectome was computed by computing the cross covariance between low passed filtered membrane potentials time courses of pairs of neurons. The used filter transfer function was $H(s)=\frac{1}{30s+1}$. The max of the absolute value of the cross covariance was extracted and used to compute the functional connectome link weights.

%\bibliography{BioCyb_MBIGD}

\end{document}